Nanostructured epoxies based on the self-assembly of block copolymers: a new miscible block that can be tailored to different epoxy formulations

**S. Maiez-Tribut,[†] J. P. Pascault,[\*,†] E. R. Soulé,[‡] J. Borrajo,[‡] and R. J. J. Williams[‡]**

*Laboratoire des Matériaux Macromoléculaires (LMM)/Ingénierie des Matériaux Polymères (IMP), UMR 5627 CNRS-INSA Lyon, Bât. Jules Verne, 20 Av. Albert Einstein, 69621 Villeurbanne Cedex, France; Institute of Materials Science and Technology (INTEMA), University of Mar del Plata and National Research Council (CONICET), Av. J. B. Justo 4302, 7600 Mar del Plata, Argentina*

\* Corresponding author. E-mail: jean-pierre.pascault@insa-lyon.fr

[†] LMM/IMP

[‡] INTEMA

ABSTRACT: Nanostructured thermosets may be obtained by the self-assembly of amphiphilic block copolymers (BCP) in a reactive solvent and fixation of the resulting morphologies by the cross-linking reaction. Nanostructuration requires the presence of a bock that remains miscible in the thermosetting polymer during polymerization. The selection of the miscible block depends on the particular system and in some cases (e.g., for epoxy-amine network based on diglycidyl ether of bisphenol A, DGEBA, and 4,4'-diaminodiphenylsulfone, DDS) it is very difficult to find such a block. In this manuscript it is shown that random copolymers of methyl methacrylate (MMA) and *N*,*N*-dimethylacrylamide (DMA) containing different molar fractions of DMA, can be used as a miscible block for the nanostructuration of epoxies, a fact that is particularly illustrated for a DGEBA-DDS epoxy network. The miscibility of the random copolymer during formation of the epoxy network was first analyzed determining cloud-point conversions as a function of the molar fraction of DMA in the copolymer. A thermodynamic model of the phase separation was performed using the Flory-Huggins model and taking the polydispersities of both polymers into account. A single expression of the interaction parameter based on the theory of random copolymers provided a reasonable fitting of the experimental cloud-point curves. The significant increase in the miscibility produced by using small DMA molar fractions in the copolymer was explained by the high negative value of the binary interaction energy between DMA and the epoxy-amine solvent, associated to the positive value of the interaction energy between DMA and MMA units. Block copolymers with poly(n-butyl acrylate) as the immiscible block and the random copolymer P(MMA-*co*-DMA) as the miscible block were used for the nanostructuration of DGEBA-DDS networks. The necessary molar fraction of DMA in the miscible block to stabilize a dispersion of nanosize domains depended on the fraction of the immiscible block in the BCP.

**Introduction**

Nanostructured thermosets may be obtained by the self-assembly of amphiphilic block copolymers (BCP) in a reactive solvent and fixation of the resulting morphologies by the cross-linking reaction.[1] In particular, BCP self-assembled into vesicles and micelles can significantly increase the fracture resistance of cured epoxies with a minimum impact on glass transition temperature and modulus.[2-6] This has important implications for the manufacture of printed circuit boards, composites and other applications.[4]

Diblock copolymers used for these purposes are composed of one block that is immiscible in the thermoset precursors and another one that is initially miscible and does not phase separate during the network formation at least up to very high conversions. In this way the self-assembled structure is fixed by the cross-linking reaction.[1,7,8] Another possibility of generating self-assembled structures is to start with a diblock copolymer with both blocks being initially miscible in the reactive solvent. Phase separation of one of the blocks induced by polymerization may also lead to a nanostructured thermoset if the other block remains miscible in the reactive solvent up to high conversions.[9] Tri- and tetrablock copolymers have been also employed with at least one block exhibiting a high miscibility during polymerization.[5,6,10,11]

Various immiscible blocks have been employed to generate stable nanostructures in epoxies based on diglycidyl ether of bisphenol A (DGEBA) cured with different hardeners. Examples are: poly(ethyl ethylene),[7] poly(ethylene-*alt*-propylene),[2-4,7,8] polyisoprene,[2] poly(styrene-*b*-butadiene),[5,6,10,11] poly(propylene oxide),[12,13] polybutadiene,[3,9] poly(2-ethylhexylmethacrylate),[3,4] and polyethylene.[14]

The election of the miscible block is strongly dependent on the hardener selected to perform the cure. Both inert and reactive miscible blocks have been reported for specific DGEBA-hardener combinations. While the former keep their miscibility up to high conversions due to the chemical compatibility with the components of the epoxy formulation (e.g., presence of specific interactions), the latter bear functional groups that react with one or both monomers preventing a macroscopic phase separation of the miscible block and allowing for the covalent bonding of the block copolymer with the epoxy matrix. Examples of inert miscible blocks are: poly (ethylene oxide),[2-4,7,8,12-14] poly(methyl methacrylate),[3,5,10,11] and poly($\varepsilon$-caprolactone).[9] Examples of miscible blocks bearing reactive groups are: epoxidized polyisoprene,[3] poly(glycidyl methacrylate),[11] poly(methyl methacrylate-*co*-glycidyl methacrylate),[3,4] and poly(methyl methacrylate-*co*-methacrylic acid).[6]

The search of a miscible block for a specific DGEBA-hardener combination is not a trivial task due to the variety of mechanisms of network formation involving different types of hardeners and the fact that commercial formulations frequently contain other epoxy monomers apart from DGEBA (e.g., brominated DGEBA for flame retardation). Poly(methyl methacrylate) (PMMA) may be a convenient selection as a miscible block because it is soluble with DGEBA in all proportions.[15-18] However, for most hardeners it becomes phase separated during polymerization well before gelation.[15,17-22] On the other hand, poly(*N*,*N*-dimethylacrylamide) (PDMA) is miscible both in non-polar solvents such as cyclohexane and in highly-polar solvents such as water, methanol and ethanol.[23] Therefore, the family of random copolymers poly(MMA-*co*-DMA), with different proportions of both monomers, should be a useful choice as a universal miscible block for the synthesis of nanostructured epoxies.

The polymerization-induced phase separation of blends of the random copolymer in a reactive solvent based on DGEBA and 4,4'-diaminodiphenylsulfone (DDS) as hardener will be first analyzed. DDS was chosen because it is one of the most often used hardeners in composites. Nanostructured networks based on DGEBA-DDS have been synthesized using reactive block copolymers[6] but there are no reported results based on the use of a miscible block. The effect of varying the molar fraction of DMA in the random copolymer on the cloud-point conversion will be assessed. A thermodynamic analysis of miscibility during polymerization will enable us to discuss the values of the binary interaction energies among the different constitutional repeating units: MMA-reactive solvent; DMA-reactive solvent and MMA-DMA, and the influence they have on the miscibility of the random copolymer. These random copolymers were then used as the miscible hard block in BCPs synthesized using poly(n-butyl acrylate) as the immiscible soft block, to generate and stabilize a dispersion of nanosize domains in the DGEBA-DDS epoxy network. A complete analysis of the different types of nanostructures that may be obtained and the properties of the resulting networks will be the subject of another publication.

**Experimental Section**

**Materials.** Table 1 shows molar masses, solubility parameters and densities of the epoxy monomer based on diglycidylether of bisphenol A (DGEBA), the hardener (4,4'-diaminodiphenylsulfone, DDS), and the random copolymers of methyl methacrylate (MMA) and dimethylacrylamide (DMA), with same range of molar masses and different contents of DMA determined by $^1$H NMR.

Tables 2 and 3 show molar masses and compositions of diblock and triblock copolymers that were tested to prove the ability of the random poly(MMA-co-DMA) block as the miscible block of the amphiphilic block copolymers used to generate nanostructured phases in a DGEBA-DDS epoxy system. The immiscible block was poly(n-butyl acrylate), PBA.

Both random and block copolymers were prepared by a Nitroxide Mediated Polymerization (NMP). The controlled free radical polymerization was performed using the alkoxyamines based on *N*-ter-butyl-*N*-(1-diethylphosphono-2,2-dimethylpropyl) nitroxide (DENP), SG1. For random copolymers, polymerizations were carried out in 1,4-dioxane at 100 ºC. To synthesize BCPs, in a first step the mono and difunctional PBA macroinitiators were prepared by bulk polymerization of butyl acrylate with the alkoxyamines, mono or difunctionnal, based on SG1: methyl methacrylic acid-SG1 MAMA-SG1 or DIAMS at 120°C. In the second step, these PBA macroinitiators were used to initiate the copolymerization of MMA and DMA in 1,4- dioxane at 100°C. The composition of copolymers and conversions of monomers were determined by $^1$H NMR in CDCl$_3$.[25]

**Cloud-point conversions.** Blends of a particular copolymer in thermoset precursors were prepared by first dissolving the copolymer in DGEBA at 135 ºC and then adding the stoichiometric amount of DDS while stirring, until a homogeneous solution was obtained. The reaction was carried out in a test tube kept at the desired temperature (110 ºC, 135 ºC or 160 ºC), inside a light transmission device. The cloud-point time was determined as the onset of the decrease of the intensity of light transmitted through the sample that was continuously recorded with a photodetector. The corresponding conversion was determined by rapidly cooling the test tube at the cloud point, dissolving its contents in a pre-determined amount of THF and determining the residual amount of DGEBA by size exclusion chromatography. Conversion is defined as:[26]

$$p = 1 - [(DGEBA)/(DGEBA)_0]^{1/2} \qquad (1)$$

where (DGEBA)$_0$ is the DGEBA concentration in the initial blend.

**Transmission electron microscopy (TEM).** TEM images were obtained with a Philips CM120 microscope applying an acceleration voltage of 80 kV. The specimens were prepared using an ultramicrotome. Thin sections of about 70 nm were obtained with a diamond knife at room temperature and deposited on copper grids. Two methods were used to stain samples: (i) the staining was performed by laying down the samples on the top of a solution containing 2 wt% of phosphotungstic acid and 2 wt% of benzyl alcohol, during 5 min at 60 °C. Then they were rinsed several times with water and were dried off with air. (ii) The samples were stained 5 min at room temperature with RuO$_4$ vapors.

**Results and discussion**

**Thermodynamic analysis of the miscibility of the random copolymer in DGEBA-DDS.** A thermodynamic analysis of the miscibility of the random copolymer in the thermoset in the course of polymerization can be carried out using the experimental values of phase-separation conversions obtained for different compositions and polymerization temperatures. This assessment has the implicit assumption that phase separation occurs when the blend enters the metastable region of the phase diagram, at a experimental conversion that should be higher but very close to the thermodynamic cloud-point conversion. The obvious condition to fulfil this requirement is that the phase separation rate should be faster than the polymerization rate. Experimental proofs of polymerization-induced phase separations proceeding very close to the thermodynamic cloud-point conversion have been reported for blends of a methacrylic monomer undergoing a free-radical polymerization in the presence of a modifier.[27-29] In these systems phase-separation conversions determined in physical blends of the monomer, the linear polymer and the modifier were the same, within experimental error, to those determined in situ during polymerization employing usual initiator concentrations and temperatures.

For the DGEBA-DDS system we selected three polymerization temperatures: 110 ºC, 135 ºC, and 160 ºC, where the reaction rates were, respectively, very slow, slow and slow/moderate. Even at the highest of these temperatures the polymerization rate was slow enough to enable an accurate determination of the conversion at the time where phase separation was observed. Under these conditions it will be assumed that experimental values of conversions at the onset of phase separation agree with thermodynamic cloud-point conversions. In the selected temperature range the only significant reaction is the stepwise polymerization involving epoxy and amine hydrogens. The polyetherification of epoxy groups is only significant at higher temperatures and at high conversions of amine hydrogens.[30] Therefore, there is no change of the reaction mechanism with temperature in the selected range. The only idealization that will be included in the thermodynamic model is that the distribution of epoxy-amine species as a function of conversion may be calculated assuming equal reactivity of primary and secondary amine hydrogens.

Based on the values of solubility parameters reported in Table 1 it may be inferred that PDMA should have a higher solubility in DGEBA-DDS than PMMA, a fact that was confirmed by cloud-point experiments. Figure 1 shows experimental values of cloud-point conversions for blends containing random copolymers of different compositions, polymerized at 135 ºC. A very small amount of DMA in the copolymer led to a significant increase in miscibility evidenced by the shift of the cloud-point conversion to higher values. The range of copolymer compositions was selected to produce phase separation before gelation in order to obtain domains with characteristic sizes measurable by the scattering of visible light (experimental gel conversion close to 0.60). Increasing further the DMA content in the copolymer shifted the cloud-point conversions to the postgel stage; a critical DMA amount will keep miscibility up to the end of reaction. Therefore these random copolymers might be used as the miscible block of amphiphilic block copolymers used to generate nanostructed phases in a variety of epoxy networks. The necessary molar fraction of DMA in the copolymer should be determined for any specific formulation.

Figure 2 shows the influence of the polymerization temperature on the cloud-point conversions for one particular random copolymer. A lower-critical-solution-temperature (LCST) behavior was observed, evidenced by the decrease in the cloud-point conversion when increasing the reaction temperature. This is probably related to the presence of specific interactions between the copolymer and epoxy-amine species with an equilibrium constant that decreases when increasing temperature. A similar lower-critical-solution-temperature (LCST) behavior was observed for the polymerization-induced phase separation of solutions of PMMA in DGEBA/DDS (Figure 3). LCST behaviors have been reported previously in the case of Polyether sulfone / DGEBA-DDS blends, [31] or for polyethylene oxide dissolved in DGEBA-Methylene Dianiline. [32]

The Flory-Huggins model was used to fit the experimental curves. The free energy per mol of unit cells with molar volume $V_r$, may be written as:

$$\Delta G/RT = \Sigma\Sigma(\phi_{m,n}/r_{m,n})\ln\phi_{m,n} + \Sigma(\phi_{2i}/r_{2i})\ln\phi_{2i} + \chi\phi_1\phi_2 \qquad (2)$$

where $\phi_{m,n}$ represents the volume fraction of species of the thermosetting polymer with $m$ units of the diamine and $n$ units of the diepoxide ($\Sigma\Sigma\phi_{m,n} = \phi_1$), $r_{m,n}$ is the molar volume of this generic species measured with respect to the reference volume, $V_r$ (taken as the molar volume of DDS = 186.5 cm$^3$/mol, approximating the density of amorphous DDS by the value of the crystalline phase reported in Table 1), $\phi_{2i}$ represents the volume fraction of the *i*-mer of the copolymer ($\Sigma\phi_{2i} = \phi_2$), $r_{2i}$ is the molar volume of the *i*-mer measured with respect to the reference volume, and $\chi$ is the interaction parameter.

The distribution of species of the copolymer was obtained from the average values of molar masses assuming a Schulz-Zimm distribution.[33] The distribution of species of the thermosetting polymer at any conversion (*p*) was obtained assuming an ideal stepwise polymerization:[34]

$$N_{m,n} = 4A_0(3m)!p^{m+n-1}(1-p)^{2m+2}/[(n-m+1)!(3m-n+1)!m!] \tag{3}$$

where $N_{m,n}$ is the molar concentration of a generic species containing *m* diamine units and *n* diepoxide units, and $A_0$ is the initial molar concentration of the diamine in the stoichiometric mixture (calculated assuming that there was no volume variation upon mixing).

The volume fraction of the generic epoxy-amine species is given by:

$$\phi_{m,n} = (N_{m,n}\, r_{m,n} / \Sigma\Sigma\, N_{m,n}\, r_{m,n})\phi_1 \tag{4}$$

where

$$r_{m,n} = (m248/1.33 + n382/1.15)/V_r \tag{5}$$

Distributions were truncated including a number of species necessary to obtain the experimental value of $M_w$ for the copolymer and the ideal theoretical value for the thermosetting polymer,[34] with a deviation less than 0.1 %.

The interaction parameter was taken as a typical function of temperature,

$$\chi = a + b/T \qquad (6)$$

where the factor $b$ that incorporates the total number of binary interactions, should be a negative value to simulate the experimental LCST behavior. For the particular case of the interactions among the units of a random copolymer (MMA and DMA) and a solvent (epoxy-amine, E), the factor $b$ is given by:[35]

$$b = (V_r/R)[B_{MMA-E} + (B_{DMA-E} - B_{MMA-E} - B_{MMA-DMA})\phi_{DMA} + B_{MMA-DMA}\phi_{DMA}^2] \qquad (7)$$

where $\phi_{DMA}$ is the volume fraction of DMA in the random copolymer and $B_{ij}$ the interaction energy per unit volume of the couple i-j. An implicit assumption of eq 7 is that the quality of the solvent (E) is the same in the conversion range of the experimental cloud-point conversions. One of the required values of interaction energies was previously reported:[36] $B_{MMA-DMA} = + 11.6$ J/cm$^3$. The two remaining values were taken as fitting parameters using the procedure described below.

Chemical potentials for both components (*1*: thermoset and *2*: random copolymer) were derived from eq 2 using standard procedures.[37] Equating them in both phases led to a couple of algebraic equations written in terms of two separation factors that relate the concentrations of every species in the initial and the emergent phases.[37] Both separation factors are related by a third equation stating that the summation of volume fractions of all species in the emergent phase equals 1. Therefore, the final system consists of three algebraic equations in three unknowns: two separation factors and the interaction parameter. These equations were solved for every experimental point using Mathcad 2001 Professional. This led to a series of values of the interaction parameter for different temperatures, initial compositions and particular random copolymers.

The values of interaction parameters obtained for solutions of random copolymers in the thermosetting polymer were correlated using eqs 6 and 7 to fit the values of $a$, $B_{MMA-E}$ and $B_{DMA-E}$. Resulting values were $a = 0.325$, $B_{DMA-E} = -19.4$ J/cm$^3$ and $B_{MMA-E} = -3.3$ J/cm$^3$. Continuous curves shown in Figures 1 to 3 correspond to the fitting obtained with these values. The fitting is reasonable taking into account the hypotheses used in the derivation of the thermodynamic model (ideal polymerization, interaction parameter independent of conversion and concentration, constant value of $a$ for the different copolymers).

Interaction parameters for the couples PMMA-Thermoset and PDMA-Thermoset, at 135 ºC, are: $\chi$(PMMA-Thermoset) = 0.142 and $\chi$(PDMA-Thermoset) = - 0.739. Therefore, PMMA becomes phase separated in the course of polymerization while PDMA remains completely miscible in the DGEBA-DDS network. The high negative value observed for $B_{DMA-E}$ is explained by the strong hydrogen bonds between the constitutional repeating units of both components. The high solubility of random copolymers containing DMA is explained by both the high negative value of the interaction energy of DMA with the solvent and the repulsion between DMA and MMA units. The latter acts favoring the miscibility of the random copolymer as it decreases the value of the resulting interaction parameter.[35]

**Miscibility of the random copolymer.** The interaction parameter between the random copolymer and DGEBA-DDS depends both on temperature and on DMA content:

$$\chi = 0.325 + \frac{V_r}{RT}\left[11.6\Phi_{DMA}^2 - 27.7\Phi_{DMA} - 3.3\right] \qquad (8)$$

By the use of eq 8 and transforming volume fractions into molar fractions, $\chi$ can be plotted as a function of the molar fraction of DMA in the random copolymer for different temperatures. Figure 4 shows clearly that for a given $\chi$ value the DMA content has to be increased when increasing the polymerization temperature, reflecting the LCST behavior.

The minimum DMA content for having a miscible random block may be estimated by making $\chi = 0$. This minimum increases with temperature as shown in Figure 4. Required molar fractions of DMA are 9 % at 110°C, 10 % at 135°C and 12 % at 160°C.

**Behavior of Block Copolymers.** To prove the ability of the random poly(MMA-*co*-DMA) block to behave as the miscible block of amphiphilic BCPs used to generate nanostructed phases in a DGEBA-DDS epoxy system, di- and triblock copolymers with a poly(n-butyl acrylate), PBA, immiscible block were synthesized (Table 2). The BCPs differed in the structure (di- or triblock) and in the DMA contents of the random block but had similar contents of PBA. Solutions of the epoxy precursors containing 5 wt % of a particular BCP were polymerized at 135°C. Table 2 gives the conversion at the cloud point, $p_{cp}$, when a reaction induced phase separation took place.

The first observation arising from Table 2 is that the molar fraction of DMA must be increased with respect to the value estimated from Figure 4 to avoid phase separation when the random copolymer is used as one of the blocks of the BCP. While a molar fraction of 10 % DMA was sufficient to avoid phase separation for the polymerization of solutions of the random copolymer in the epoxy precursors at 135 °C, a similar composition led to phase separation before gelation in the case of the block copolymer. This may be qualitatively explained by the fact that the aggregation of the immiscible blocks confines the miscible blocks in the same region of space. This produces a decrease in both the absolute value of the entropic contribution to the free energy and in the local concentration of solvent in contact with chains of the random copolymer (a fraction of solvent-chain interactions is replaced by chain-chain interactions). Both factors produce a decrease in the miscibility of the random copolymer when it becomes a block of the BCP. It might also be argued that the observed larger immiscibility of the random block when it is joined to an immiscible block in a block copolymer is produced by an increase in the phase separation rate rather than by thermodynamic arguments. However, as the polymerization rate is slow at the selected temperatures, relative changes in phase separation rates should not affect the experimental values of conversions at the onset of phase separation.

For the three phase-separated formulations shown in Table 2, the size of dispersed domains increased beyond the nanometer range (opaque materials were obtained). Increasing the molar fraction of DMA units in the random block led to an increase in the cloud-point conversion as expected. When the molar fraction of DMA in the random block was increased to 25 % samples conserved the transparency up to the end of polymerization at 135°C. This is clearly illustrated by TEM micrographs shown in Figure 5. The opaque samples exhibit relatively large dispersed domains (Figures 5a and 5b). Samples containing high molar fractions of DMA in the random block exhibit dispersed nanoparticles with diameters in the order of 25nm (Figures 5c and 5d).

In order to analyze the effect of increasing amounts of the PBA immiscible block when keeping constant the molar fraction of DMA in the random block, BCPs containing a molar fraction of DMA close to 8 % in the random block and variable amounts of the PBA block, were synthesized (Table 3). The selected molar fraction of DMA units enabled to observe phase separation in every formulation for polymerizations carried out at 135 °C. In Figure 6 conversions at the cloud point, $p_{cp,}$ for blends with 5 wt % diblocks (from Table 3) are plotted versus the PBA concentration. Increasing the fraction of PBA in the block copolymer produced a decrease in the miscibility of the random block (decrease in the cloud-point conversion). This may be explained by the increase in the average size of the immiscible PBA domains before polymerization. This will confine more miscible chains of the random block in the same region of space decreasing the entropic contribution and replacing solvent-chain interactions by chain-chain interactions. Both factors produce a decrease in the miscibility of the block constituted by the random copolymer. Therefore, the molar fraction of DMA units in the random copolymer that is necessary to avoid phase separation and stabilize the nanostructure at the end of the reaction depends also on the fraction of PBA in the BCP.

Figure 7 shows TEM micrographs obtained for a fully cured sample prepared with 10 wt % of triblock copolymer containing a molar fraction of 33 % BA units and a molar fraction of 25 % DMA units in the random P(MMA-co-DMA) blocks. The initial transparent solution was polymerized during 20 hours at 135°C and then postcured 6 hours at 220°C. Well dispersed nanoparticles with PBA cores and diameters of 20-30 nm are present in the epoxy matrix. It means that if enough DMA units are introduced in the random block, nanostructured thermosets based on a DGEBA-DDS epoxy system can be synthesized.

## 4. Conclusions

Random copolymers of MMA and DMA may be employed as the miscible block in amphiphilic block copolymers used to generate nanostructured phases in epoxy networks. Increasing the amount of DMA in the copolymer increases miscibility due to the strong specific interactions between DMA and the epoxy-amine solvent combined with the repulsion between DMA and MMA units. The possibility of varying the DMA fraction in the MMA-DMA random block gives the possibility to tailor these new BCPs for their use in a variety of epoxy-hardener formulations.[38] This was illustrated by using these random copolymers as the miscible block of BCPs with PBA as the immiscible block. The BCPs were used to generate and stabilize a dispersion of nanosize PBA domains in epoxy networks based on DGEBA-DDS.

**Acknowledgments.** French authors acknowledge the financial support of Arkema and CNRS. Argentine authors acknowledge the financial support of the University of Mar del Plata, ANPCyT (PICT03 14738) and CONICET, Argentina. Prof. D.Bertin and Dr Trang N.T. Phan (CROPS - UMR 6517 - Université de Provence, Marseille, France) are acknowledged for the synthesis of random and block copolymers. The support of the European Network of Excellence Nanofun-Poly for the diffusion of these results is also acknowledged.

**References and Notes**

(1) Pascault, J. P.; Williams, R. J. J. In *Micro- and nanostructured multiphase polymer blend systems*; Harrats, C.; Thomas, S.; Groeninckx, G., Eds.; CRC Press-Taylor and Francis: Boca Raton (FL), 2006; p. 359-390.

(2) Dean, J. M.; Lipic, P. M.; Grubbs, R. B.; Cook, R. F.; Bates, F. S. *J. Polym. Sci., Part B: Polym. Phys.* **2001**, *39*, 2996.

(3) Dean, J. M.; Grubbs, R. B.; Saad, W.; Cook, R. F.; Bates, F. S. *J. Polym. Sci., Part B: Polym. Phys.* **2003**, *41*, 2444.

(4) Dean, J. M.; Verghese, N. E.; Pham, H. Q.; Bates, F. S. *Macromolecules* **2003**, *36*, 9267.

(5) Ritzenthaler, S.; Court, F.; Girard-Reydet, E.; Leibler, L.; Pascault, J. P. *Macromolecules* **2003**, *36*, 118.

(6) Rebizant, V.; Venet, A. S.; Tournilhac, F.; Girard-Reydet, E.; Navarro, C.; Pascault, J. P.; Leibler, L. *Macromolecules* **2004**, *37*, 8017.

(7) Hillmeyer, M. A.; Lipic, P. M.; Hajduk, D. A.; Almdal, K.; Bates, F. S. *J. Am. Chem. Soc.* **1997**, *119*, 2749

(8) Lipic, P. M.; Bates, F. S.; Hillmeyer, M. A. *J. Am. Chem. Soc.* **1998**, *120*, 8963.

(9) Meng, F.; Zheng, S.; Zhang, W.; Li, H.; Liang, Q. *Macromolecules* **2006**, *39*, 711.

(10) Ritzenthaler, S.; Court, F.; David, L.; Girard-Reydet, E.; Leibler, L.; Pascault, J. P. *Macromolecules* **2002**, *35*, 6245.

(11) Rebizant, V.; Abetz, V.; Tournilhac, F.; Court, F.; Leibler, L. *Macromolecules* **2003**, *36,* 9889.

(12) Mijovic, J.; Shen, M.; Sy, J. W.; Mondragon, I. *Macromolecules* **2000**, *33*, 5235.

(13) Guo, Q.; Thomann, R.; Gronski, W.; Thurn-Albrecht, T. *Macromolecules* **2002**, *35*,

3133.

(14) Guo, Q.; Thomann, R.; Gronski, W.; Staneva, R.; Ivanova, R.; Stühn, B. *Macromolecules* **2003**, *36*, 3635.

(15) Gómez, C. M.; Bucknall, C. B. *Polymer* **1993**, *34*, 2111.

(16) Woo, E. M.; Wu, M. N. *Polymer* **1996**, *37*, 2485.

(17) Ritzenthaler, S.; Girard-Reydet, E.; Pascault, J. P. *Polymer* **2000**, *41*, 6375.

(18) González García, F.; Soares, B. G.; Williams, R. J. J. *Polym. Int.* **2002**, *51*, 1340.

(19) Mondragon, I.; Remiro, P. M.; Martin, M. D.; Valea, A.; Franco, M.; Bellenguer, V. *Polym. Int.* **1998**, *47*, 152.

(20) Galante, M. J.; Oyanguren, P. A.; Andromaque, K.; Frontini, P. M.; Williams, R. J. J. *Polym. Int.* **1999**, *48*, 642.

(21) Remiro, P. M.; Riccardi, C. C.; Corcuera, M. A.; Mondragon, I. *J. Appl. Polym. Sci.* **1999**, *74*, 772.

(22) Galante, M. J.; Borrajo, J.; Williams, R. J. J.; Girard-Reydet, E.; Pascault, J. P. *Macromolecules* **2001**, *34*, 2686.

(23) El-Ejmi, A. A. S; Huglin, M. B. *Polym. Int.* **1996**, *39*, 113.

(24) Van Krevelen, D. W. *Properties of Polymers*; 3$^{rd}$ Edition; Elsevier: Amsterdam, 1990.

(25) Trang, N. T. P.; Bertin, D.; Maiez-Tribut, S.; Pascault, J. P.; Gerard, P.; Guerret, O. to be submitted.

(26) Verchère, D.; Sautereau, H.; Pascault, J. P.; Riccardi, C. C.; Moschiar, S. M.; Williams, R. J. J. *Macromolecules* **1990**, *23*, 725.

(27) Liskova, A.; Berghmans, H. *J. Appl. Polym. Sci.* **2004**, *91*, 2234.

(28) Schnell, M.; Borrajo, J.; Williams, R. J. J.; Wolf, B. A. *Macromol. Mater. Eng.* **2004**, *289*, 642.

(29) Soulé, E. R.; Borrajo, J.; Williams, R. J. J. *Macromolecules* **2005**, *38*, 5987.

(30) Riccardi, C. C.; Williams, R. J. J. *J. Appl. Polym. Sci.* **1986**, *32*, 3445.

(31) Yamanaka, K.; Inoue, T. *Polymer,* **1989**, *30, 662*

(32) Larranaga, M.; Gabilondo, N.; Kortaberria, G.; Serrano, E.; Remiro, P.; Riccardi, C.C.; Mondragon, I.; *Polymer,* **2005**, *46, 7082*

(33) Pascault, J. P.; Williams, R. J. J. In *Polymer Blends, Vol. 1*; Paul, D. R.; Bucknall, C. B.; Eds.; Wiley-Interscience: New York, 2000; p. 379-415.

(34) Peebles, L. H. *Molecular Weights Distributions in Polymers*; Wiley-Interscience: New York, 1971.

(35) Merfeld, G. D.; Paul, D. R. In *Polymer Blends, Vol. 1*; Paul, D. R.; Bucknall, C. B.; Eds.; Wiley-Interscience: New York, 2000; p. 55-91.

(36) Soulé, E. R.; Jaffrennou, B.; Méchin, F.; Pascault, J. P.; Borrajo, J.; Williams, R. J. J. *J. Polym. Sci., Part B: Phys.* **2006**,44, 2821.

(37) Kamide, K. *Thermodynamics of Polymer Solutions: Phase Equilibria and Critical Phenomena*; Elsevier: Amsterdam, 1990.

(38) Guerret, O.; Gerard, P.; Pascault, J. P.; Bonnet, A. Patent PCT/EP 2006/00564.

| Product | $\delta\,(\text{J/cm}^3)^{1/2}$ | DMA molar% | $M_n$ (g/mol) | $M_w$ (g/mol) | $\rho\,(\text{g/cm}^3)$ |
| --- | --- | --- | --- | --- | --- |
| DGEBA | 20.7 | | 382 | | 1.15 |
| DDS | 23.5 | | 248 | | 1.33 |
| PMMA | 18.7 | 0 | 15000 | 21000 | 1.20 |
| P(MMA-co-DMA)3.5 | | 3.5 | 21000 | 27600 | 1.20 |
| P(MMA-co-DMA)3.9 | | 3.9 | 21000 | 27600 | 1.20 |
| P(MMA-co-DMA)5.7 | | 5.7 | 18000 | 23900 | 1.20 |
| PDMA | 22.7 | 100 | 77500 | 100 000 | 1.30 |

**Table 1. Solubility Parameters Calculated from Individual Contributions of Different Forces[24], Composition of the Random Copolymers, Average Molar Masses and Densities of Different Products**

| Diblock or triblock copolymers | PBA molar%BA | $M_n$ kg/mol PBA | DMA molar% | $M_n$ kg/mol BCP | $p_{cp}$ |
|---|---|---|---|---|---|
| P[BA-*b*-MMA] | 29 | 19 | 0 | 69 | 0.30 |
| P[BA-*b*-(MMA-*co*-DMA)] | 31 | 20 | 9 | 75 | 0.46 |
| P[(MMA-*co*-DMA)-*b*-BA-*b*-(MMA-*co*-DMA)] | 29 | 20 | 10 | 57 | 0.46 |
| P[BA-*b*-(MMA-*co*-DMA)] | 30 | 38 | 29 | 106 | No phase separation |
| P[(MMA-*co*-DMA)-*b*-BA-*b*-(MMA-*co*-DMA)] | 33 | 20 | 25 | 50 | No phase separation |

**Table 2. Effect of the wt % DMA in the Random Block on the Cloud-Point Conversion for Blends Containing 5 wt % Diblock and Triblock Copolymers in DGEBA-DDS at 135 ºC**

| Random and diblock copolymers | PBA molar%BA | $M_n$ kg/mol PBA | $M_n$ kg/mol CP | DMA molar% | $p_{cp}$ |
|---|---|---|---|---|---|
| P(MMA-*co*-DMA) | no | no | 17.5 | 7.5 | 0.62 |
| P[BA-*b*-(MMA-*co*-DMA)] | 9 | 19.6 | 180.0 | 7.5 | 0.51 |
| P[BA-*b*-(MMA-*co*-DMA)] | 12 | 12.1 | 79.5 | 9 | 0.57 |
| P[BA-*b*-(MMA-*co*-DMA)] | 20 | 19.6 | 82.7 | 9 | 0.52 |
| P[BA-*b*-(MMA-*co*-DMA)] | 26 | 22.5 | 72.4 | 9 | 0.49 |
| P[BA-*b*-(MMA-*co*-DMA)] | 31 | 20.0 | 75.1 | 9 | 0.46 |

**Table 3. Effect of the BA Fraction in the Diblock Copolymer on the Cloud Point Conversion for Blends Containing 5 wt % Diblock Copolymer in DGEBA-DDS at 135 ºC**

**Figure captions**

**Figure 1.** Cloud-point conversions measured during the polymerization at 135 ºC of blends of DGEBA-DDS and P(MMA-*co*-DMA) random copolymers containing different DMA molar fractions.

**Figure 2.** Cloud-point conversions at different polymerization temperatures for blends of the random copolymer with 3.9 molar % DMA and DGEBA-DDS.

**Figure 3.** Cloud-point conversions at different polymerization temperatures for blends of PMMA and DGEBA-DDS.

**Figure 4.** Variation of the interaction parameter $\chi$ for the blend of the random copolymer with DGEBA-DDS as a function of the molar fraction of DMA, for three different temperatures.

**Figure 5.** TEM pictures of DGEBA-DDS blends modified by 5 wt % of diblock and triblock copolymers (listed in Table 2). Samples are cured at 135°C + 6h at 220°C and stained 5 min at room temperature with $RuO_4$ vapors: a) P[BA-*b*-(MMA-*co*-DMA)] 9%DMA;  b) P[(MMA-*co*-DMA)-*b*-BA-*b*-(MMA-*co*-DMA)] 10%DMA; c) P[BA-*b*-(MMA-*co*-DMA)] 29%DMA;  d) P[(MMA-*co*-DMA)-*b*-BA-*b*-(MMA-*co*-DMA)] 25%DMA..

**Figure 6.** Cloud-point conversions for blends of the BCPs (listed in Table 3) with ~ 8 molar % DMA in DGEBA-DDS, polymerized at 135°C.

**Figure 7.** TEM micrographs of fully cured DGEBA-DDS blends with 10 wt% of P[(MMA-*co*-DMA)-*b*-BA-*b*-(MMA-*co*-DMA)]. BCP composition: 33 moles % of BA units, $M_n$(PBA) = 20 kg/mol, 25 moles % of DMA units in the random block, and $M_n$ of the triblock = 50 kg/mol. Stained with a) acid phosphotungstique + benzyl alcohol solution; b) $RuO_4$ vapors.

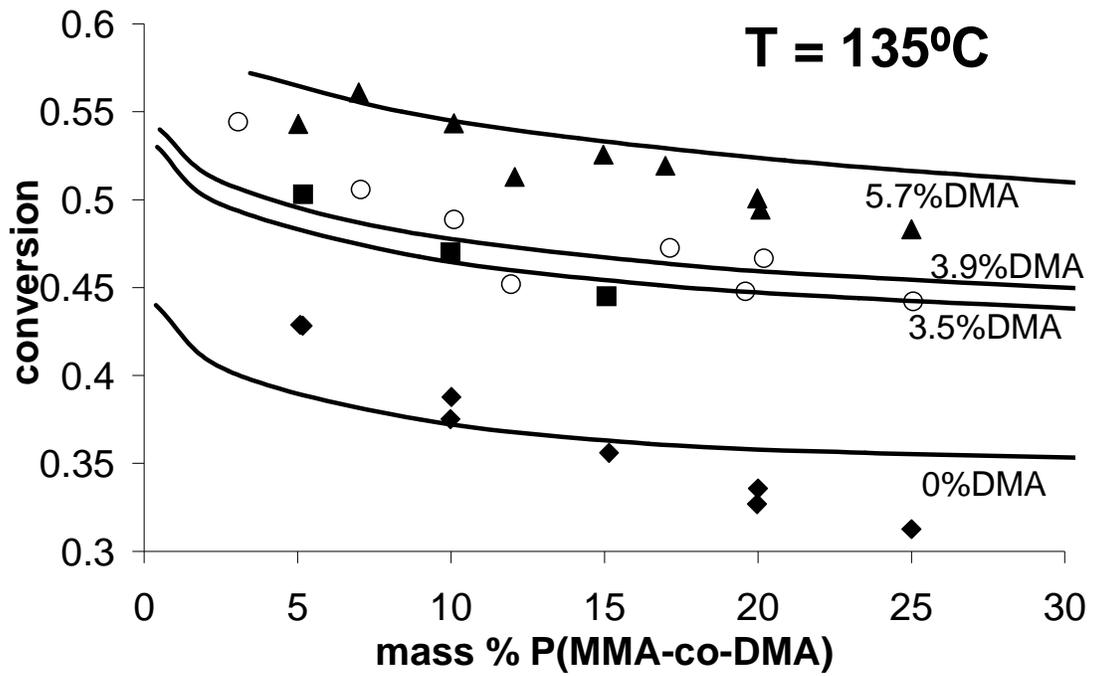

**Figure 1**

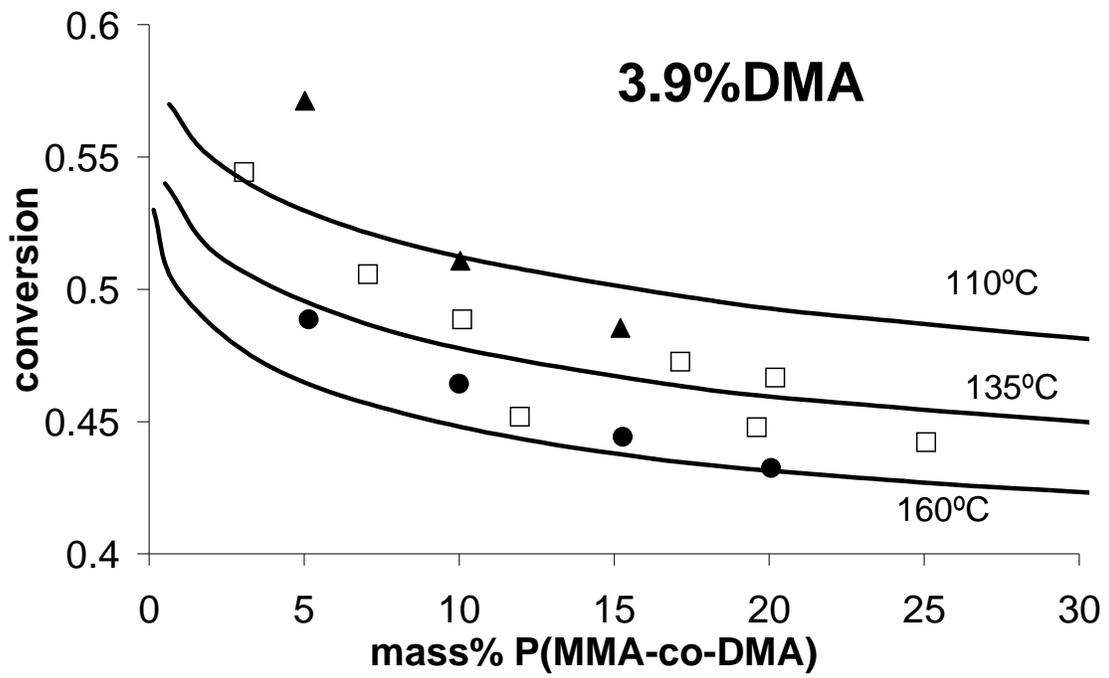

Figure 2

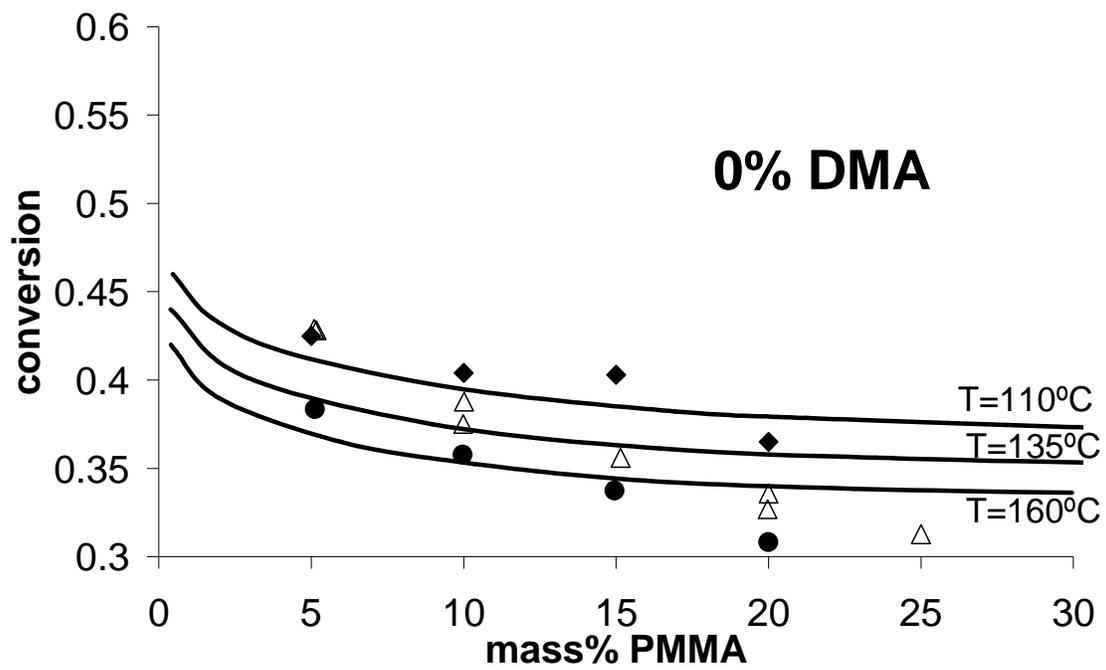

**Figure 3**

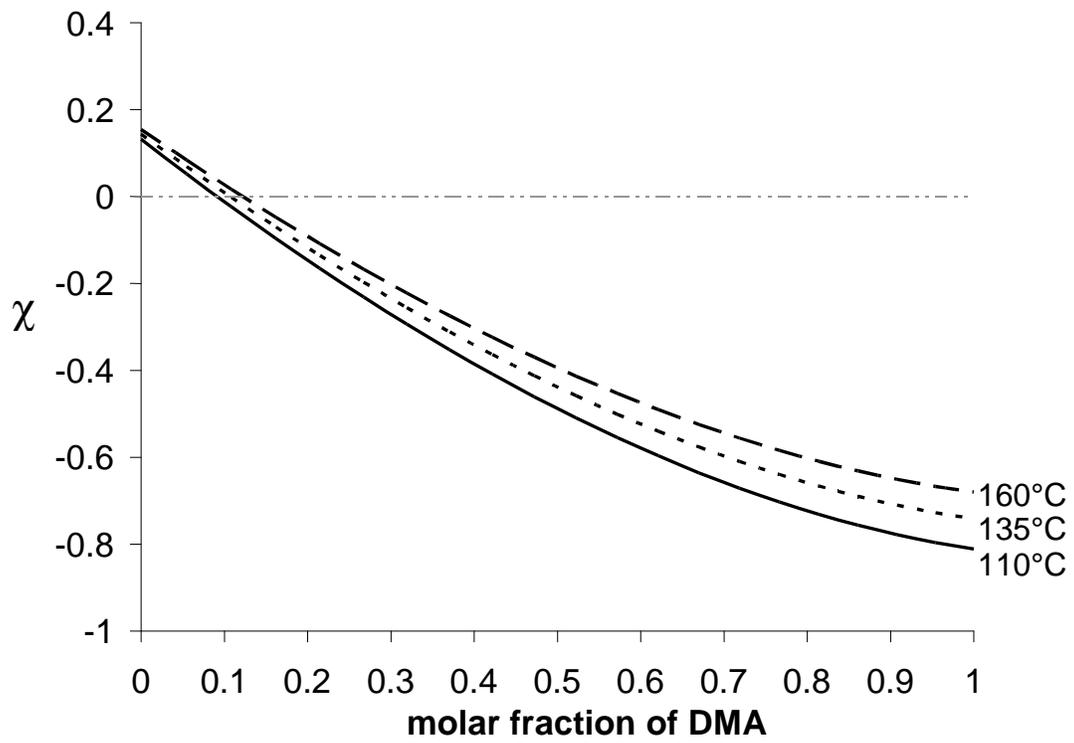

**Figure 4**

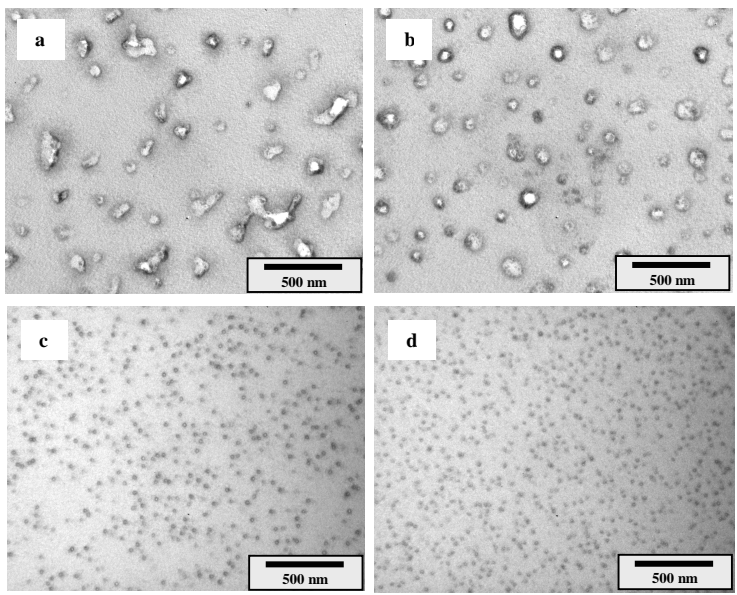

**Figure 5**

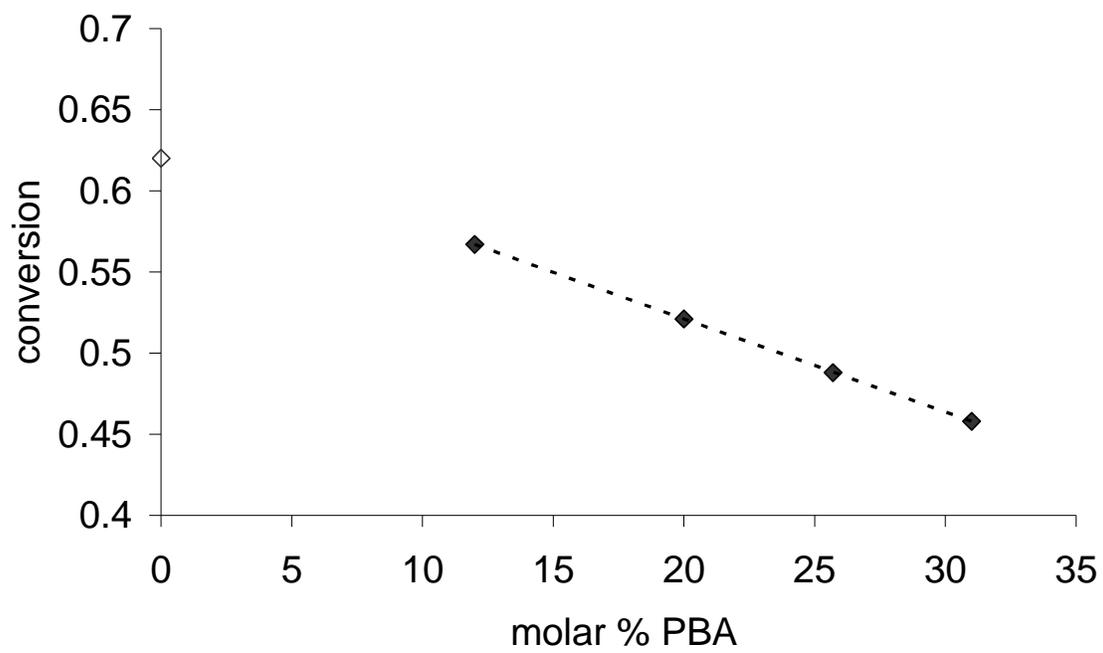

**Figure 6**

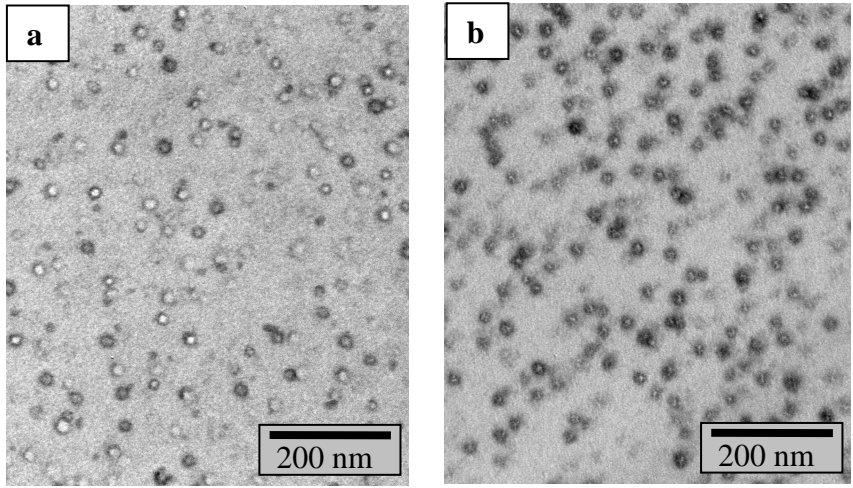

**Figure 7**